\documentclass[conference]{IEEEtran}
\IEEEoverridecommandlockouts
\usepackage{cite}
\usepackage{algorithm}
\usepackage{algpseudocode}
\usepackage{amsmath,amssymb,amsfonts}
\usepackage{graphicx}
\usepackage{xcolor}
\usepackage{todonotes}
\usepackage{booktabs}
\usepackage{textcomp}
\usepackage{xspace}
\usepackage{multirow}
\usepackage{diagbox}
\usepackage{paralist}
\usepackage{url}
\usepackage{colortbl}
\usepackage{cleveref}
\crefformat{section}{\S.#2#1#3} 
\crefformat{subsection}{\S.#2#1#3}
\crefformat{subsubsection}{\S.#2#1#3}
\def\BibTeX{{\rm B\kern-.05em{\sc i\kern-.025em b}\kern-.08em
    T\kern-.1667em\lower.7ex\hbox{E}\kern-.125emX}}

\newcommand{\vx}{\mathbf{x}}
\newcommand{\vtheta}{\mathbf{\theta}}

\newcommand{\expect}{\mathbb{E}}
\newcommand{\distrib}{\mathcal{D}}
\newcommand{\binaryset}{\{0,1\}}

\newcommand{\sgn}{\mathrm{sgn}}
\newcommand{\fgsm}{\texttt{FGSM}}
\newcommand{\mfgsm}{$\fgsm^k$}
\newcommand{\rmfgsm}{\texttt{r}\mfgsm}
\newcommand{\dmfgsm}{\texttt{d}\mfgsm}

\newcommand{\mBCA}{$\texttt{BCA}^k$}
\newcommand{\mBGA}{$\texttt{BGA}^k$}
\newcommand{\adloss}{\max_{\bar{\vx} \in \mathcal{S}(\vx)} L(\vtheta, \bar{\vx}, y)}
\newcommand{\bscn}{\mathcal{N}_{BS}}
\newcommand{\dLdx}{\nabla_{\vx}L(\vtheta, \vx^t, y)}

\newcommand{\shortdLdxj}{\partial_{x^{t}_j}L}
\newcommand{\malgan}{\texttt{MalGAN}}
\newcommand{\horse}{{\textit{\sc Sleipnir}}\xspace}
\newcommand{\AEs}{AEs\xspace}
\newcommand{\cfbox}[1]{\fcolorbox{white}{gray!30}{#1}}

\begin{document}

\title{
Adversarial Deep Learning for Robust Detection of Binary Encoded Malware
}

\author{
\IEEEauthorblockN{ Abdullah Al-Dujaili}
\IEEEauthorblockA{\textit{CSAIL}, \textit{MIT}\\
	Cambridge, USA \\
	aldujail@mit.edu}
\and
\IEEEauthorblockN{ Alex Huang}
\IEEEauthorblockA{\textit{CSAIL}, \textit{MIT}\\
	Cambridge, USA \\
	alhuang@mit.edu}
\and
\IEEEauthorblockN{Erik Hemberg}
\IEEEauthorblockA{\textit{CSAIL}, \textit{MIT}\\
Cambridge, USA \\
hembergerik@csail.mit.edu}
\and
\IEEEauthorblockN{Una-May O'Reilly}
\IEEEauthorblockA{\textit{CSAIL}, \textit{MIT}\\
Cambridge, USA \\
unamay@csail.mit.edu}
}

\maketitle

\begin{abstract}
  Malware is constantly adapting in order to avoid detection. Model
  based malware detectors, such as SVM and neural networks, are
  vulnerable to so-called \textit{adversarial
  examples} which are modest changes to detectable malware that allows the resulting malware to evade detection. 
Continuous-valued methods that are robust to adversarial examples of images have been developed using saddle-point optimization formulations.  
We are inspired by them to develop similar methods for the discrete, e.g. binary, domain which characterizes the features of malware. A specific extra challenge of malware is that the adversarial examples must be generated in a way that preserves their malicious functionality.
We
  introduce methods capable of generating functionally preserved
  adversarial malware examples in the binary domain. Using the saddle-point formulation, we incorporate the adversarial examples into the training of models that are robust to them. We evaluate the
  effectiveness of the methods and others in the literature on a set of Portable Execution~(PE)
  files. Comparison prompts our introduction of an online measure computed during training to assess general expectation of robustness. \end{abstract}


\section{Introduction}
\label{sec:introduction}

Deep neural networks~(DNN) started as extensions of neural networks in artificial
intelligence approaches to computer vision and speech
recognition. They are also used in computer security applications
such as malware detection. A large challenge in developing malware detection models is
the intelligent adversaries who actively try to evade them by judiciously perturbing the detectable malware to create 
what are called \textit{Adversarial Examples} (\AEs), i.e. malware variants that evade detection. 

Much of the work done to understand and counter \AEs has occurred in the image
classification domain. An adversarial attack on an image classifier perturbs an image so that it is
perceptually no different to a human but now classified incorrectly by the classifier. 
To counter them, researchers have demonstrated how DNN
models can be trained more robustly. These methods 
assume a continuous input domain~\cite{madry2017towards}.

Our interest is malware detection where, in contrast to images, detectors often use features represented as binary~(${0,1})$ inputs. Malware \AEs must not only fool the detector, they must also ensure that their perturbations do not alter the malicious payload. Our preliminary goal is to develop a method that, as is done in the continuous space, can generate (binary) perturbations of correctly classified malware that evade the detector. Our central goal is 
to investigate how the robust adversarial training methods
for continuous domains can be transformed to serve the discrete or categorical feature domains that include malware. We can measure the effectiveness of
a robust adversarial malware method on training a classifier by the evasion rate of
\AEs and we also seek an online training measure that expresses the general expectation of model robustness.

This leads to the following contributions at the intersection of security and adversarial machine learning: 
\begin{inparaenum}
\item We present 4 methods to generate binary-encoded \AEs of malware with preserved malicious functionality
\item We present the \horse framework for training robust adversarial malware detectors. \horse employs saddle-point optimization (hence its name\footnote{\url{https://en.wikipedia.org/wiki/Sleipnir}}) to learn malware detection models for executable files represented by binary-encoded features.
\item We demonstrate the framework on a set of Portable Executables (PEs), observing that incorporating randomization in the method is most effective.
\item We use the \AEs of an adversarial crafting method~\cite{grosse2017adversarial} that does not conform to the saddle-point formulation of \horse to evaluate the models from \horse. We find that the model of the randomized method is also robust to them.
\item Finally, we provide the \horse framework and dataset for public use.\footnote{\url{https://github.com/ALFA-group/robust-adv-malware-detection}.} 
\end{inparaenum}

The paper is structured as follows. \cref{sec:background}
presents background and related work. \cref{sec:method} describes the method. Experiments are in
\cref{sec:experiments}. Finally, conclusions are drawn and future work is outlined in 
\cref{sec:concl-future-wrok}.
\section{Background}
\label{sec:background}

Malware detection is moving away from hand-crafted rule-based
approaches and towards machine
learning techniques~\cite{schultz2001data}. In this section we focus on malware
detection with neural networks (\cref{sec:malw-detect-using}),
adversarial machine learning (\cref{sec:advers-mach-learn}) and
adversarial malware versions
(\cref{sec:adversarial-malware}).

\subsection{Malware Detection using Neural Networks}
\label{sec:malw-detect-using}

Neural network methods for malware detection are increasingly being used.
For features, one study combines DNN's with random
projections~\cite{dahl2013large} and another with two dimensional binary PE program
features~\cite{saxe2015deep}. Research has also been done
on a variety of file types, such as Android and PE
files~\cite{raff2017learning, yuan2014droid, huang2017r2,
andersonevading}. While the specifics can vary greatly, all machine learning approaches
 to malware detection share the same central vulnerability to
\AEs. 

\subsection{Adversarial Machine Learning}
\label{sec:advers-mach-learn}

Finding effective techniques that robustly handle \AEs is one focus of adversarial machine learning~\cite{huang2011adversarial, biggio2017wild}. An
{adversarial example} is created by making a small, essentially non-detectable change to 
 a data sample $\vx$ to create
$\vx_{adv} = \vx + \delta$. If the detector misclassifies
$\vx_{adv}$ despite having correctly classified $\vx$, then
$\vx_{adv}$ is a successful adversarial example. Goodfellow \textit{et al.} \cite{goodfellow2014explaining} provide a clear explanation for
the existence of \AEs. 


There are a variety of techniques that generate \AEs~\cite{goodfellow2014explaining,
szegedy2013intriguing}.  
One efficient and widely used technique is the  fast gradient sign method~(\fgsm)
\cite{goodfellow2014explaining}. With respect to an input, this method finds the
directions that move the outputs of the neural network the greatest degree
and moves the inputs along these directions by small amounts, or perturbations.
Let $\vx$ represent an input, $\mathbf{\theta}$ the
parameters of the model, $y$ the labels, and $L(\mathbf{\theta}, \vx,
y)$ be the associated loss generated by the network. Maintaining the
restriction of $\epsilon$-max perturbation, we can obtain a max-norm
output change using $\eta = \epsilon\textnormal{sgn}(\nabla_{\vx}L(\mathbf{\theta},
\vx, y))$. Because the technique references the detector's parameters,
it is known as a \textit{white-box} attack model~\cite{papernot2016limitations, goodfellow2014explaining, carlini2017adversarial}.

There have been multiple studies focused on advancing
model performance against \AEs,
e.g.~\cite{zantedeschi2017efficient,na2017cascade}. One obvious approach is retraining
with the \AEs incorporated into the training set. We are attracted to the approach of \cite{madry2017towards}. It
casts model learning as a robust optimization problem with a saddle-point
formulation where the outer minimization of detector (defensive) loss is 
tied to the inner maximization of detector loss (via \AEs)~\cite{madry2017towards}. 
The approach successfully demonstrated robustness against
adversarial images by incorporating, while training, \AEs
generated using projected gradient descent.  

\subsection{Adversarial Malware}
\label{sec:adversarial-malware}

Security researchers have generated malware \AEs using an array
of machine learning approaches such as reinforcement learning, genetic
algorithms and supervised learning including neural networks, decision trees and SVM~\cite{dang2017evading,andersonevading,grosse2016adversarial,grosse2017adversarial,hu2017generating,raff2017learning,xu2016automatically,yang2017malware,saxe2015deep}.  
These approaches, with the exception of \cite{grosse2016adversarial,grosse2017adversarial}, are black box. They assume no knowledge of the detector though the detector can be queried for detection decisions.  
Multiple studies use binary features, typically where each index acts as an indicator to express the presence or absence of an API call, e.g.~\cite{rosenberg2017generic}. One study also includes byte/entropy histogram features~\cite{saxe2015deep}. Studies to date have only retrained with \AEs.



Uniquely, this work generates functional white-box
\AEs in the discrete, binary domain while
incorporating them into the training of a malware classifier that is robust to
\AEs.

\section{Method}
\label{sec:method}
To address the problem of hardening machine learning anti-malware detectors via adversarial learning, we formulate the adversarial learning procedure as a saddle-point problem in line with~\cite{madry2017towards}. Before describing the problem formally and presenting our proposed approach to tackle the same, we introduce the notation and terminology used in the rest of the paper.

\subsection{Notation} This paper considers a malware classification task with an underlying data distribution~$\distrib$ over pairs of binary executable representations and their corresponding labels (i.e., benign or malignant). For brevity, we use \textit{malicious binary executable} and \textit{malware} interchangeably. We denote the representation space of the executables and their label space by $\mathcal{X}$ and $\mathcal{Y}$, respectively. Based on extracted static features, each binary executable is represented by a binary indicator vector $\vx=[x_1,\ldots,x_m] \in \mathcal{X}$. That is, $\mathcal{X}=\binaryset^{m}$ and $x_j$ is a binary value that indicates whether the $j$th feature is present or not. On the other hand, labels are denoted by $y \in \mathcal{Y}=\binaryset$, where $0$ and $1$ denote benign and malignant executables, respectively. We would like to learn the parameters $\vtheta \in \mathbb{R}^p$ of  a binary classifier model such that it correctly classifies samples drawn from $\distrib$.  Typically, the model's performance is measured by a scalar loss function $L(\vtheta, \vx, y)$ (e.g., the cross entropy loss). The task then is to find the optimal model parameters 
 $\vtheta^*$ that minimize the risk $\expect_{(\vx,y)\sim\distrib} [L(\vtheta, \vx, y)]$. Mathematically, we have
 \begin{equation}
 \vtheta^* \in \arg\min_{\vtheta \in \mathbb{R}^p} \expect_{(\vx,y)\sim\distrib} [L(\vtheta, \vx, y)]\;.
 \label{eq:risk}
 \end{equation}
 
\subsection{Malware Adversarial Learning as a Saddle Point Problem} 
Blind spots are regions in a model's decision space, on either side of the decision boundary, where, because no training example was provided, the decision boundary is inaccurate. Blind spots of malware detection models---such as the one learned in~\eqref{eq:risk}---can be exploited to craft misclassified adversarial  malware samples from a correctly classified malware, while still preserving malicious functionality. An adversarial malware version  $\vx_{adv}$ (which may or may not be misclassified) of a correctly classified malware $\vx$ can be generated by perturbing $\vx$ in a way that maximizes the loss $L$, i.e.,
\begin{equation}
\vx_{adv} \in \mathcal{S}^*(\vx) = \arg\max_{\bar{\vx} \in \mathcal{S}(\vx)} L(\vtheta, \bar{\vx}, y)\;,
\label{eq:adv-samples}
\end{equation}
where $\mathcal{S}(\vx) \subseteq \mathcal{X}$ is the set of binary indicator vectors that preserve the functionality of malware $\vx$, and $\mathcal{S}^*(\vx) \subseteq \mathcal{S}(\vx)$ is the set of adversarial malware versions that maximize the adversarial loss. 

To harden the model learned in~\eqref{eq:risk} against the adversarial versions generated in~\eqref{eq:adv-samples}, one needs to incorporate them into the learning process. We choose to do so by making use of the saddle-point formulation presented in~\cite{madry2017towards}. Thus, our adversarial learning
composes~\eqref{eq:risk} and~\eqref{eq:adv-samples} as:
\begin{equation}
\vtheta^* \in \underbrace{\arg\min_{\vtheta \in \mathbb{R}^p}\expect_{(\vx,y)\sim\distrib} \bigg[ \overbrace{\adloss}^\text{adversarial loss}\bigg]}_\text{adversarial learning}\;.
\label{eq:saddle-problem}
\end{equation}
Solving~\eqref{eq:saddle-problem} involves an inner non-concave maximization problem and an outer non-convex minimization problem. Nevertheless, this formulation is particularly interesting because of two reasons. First, in the case of a continuous differentiable loss function (in the model parameters $\vtheta$), Danskin's theorem states that gradients at inner maximizers correspond to descent directions for the saddle-point problem---see~\cite{madry2017towards} for a formal proof. Second, it has been shown empirically that one can still reliably optimize the saddle-point problem for learning tasks with continuous feature space---i.e., $\mathcal{X}\subseteq \mathbb{R}^m$---even with \textit{i)} loss functions that are not continuously differentiable (e.g., ReLU units); and \textit{ii)} using gradients at approximate maximizers of the inner problem~\cite{madry2017towards}.  To find these maximizers, prior  work has used variants of projected gradient descent on the negative loss function such as the Fast Gradient Sign Method (\fgsm)~\cite{goodfellow2014explaining} and its multi-step variant \mfgsm~\cite{kurakin2016adversarial}. Finding maximizers (or approximations) of the inner problem for a given malware involves moving from continuous to constrained binary optimization: flipping bits of the malware's binary feature vector $\vx$ while preserving its functionality. 


\subsection{Adapting Gradient-Based Inner Maximization Methods for Binary Feature Spaces}

\begin{figure}[b]
	\begin{tabular}{cc}
		\includegraphics[width=0.22\textwidth,trim={1cm 10.5cm 14.5cm 1cm},clip]{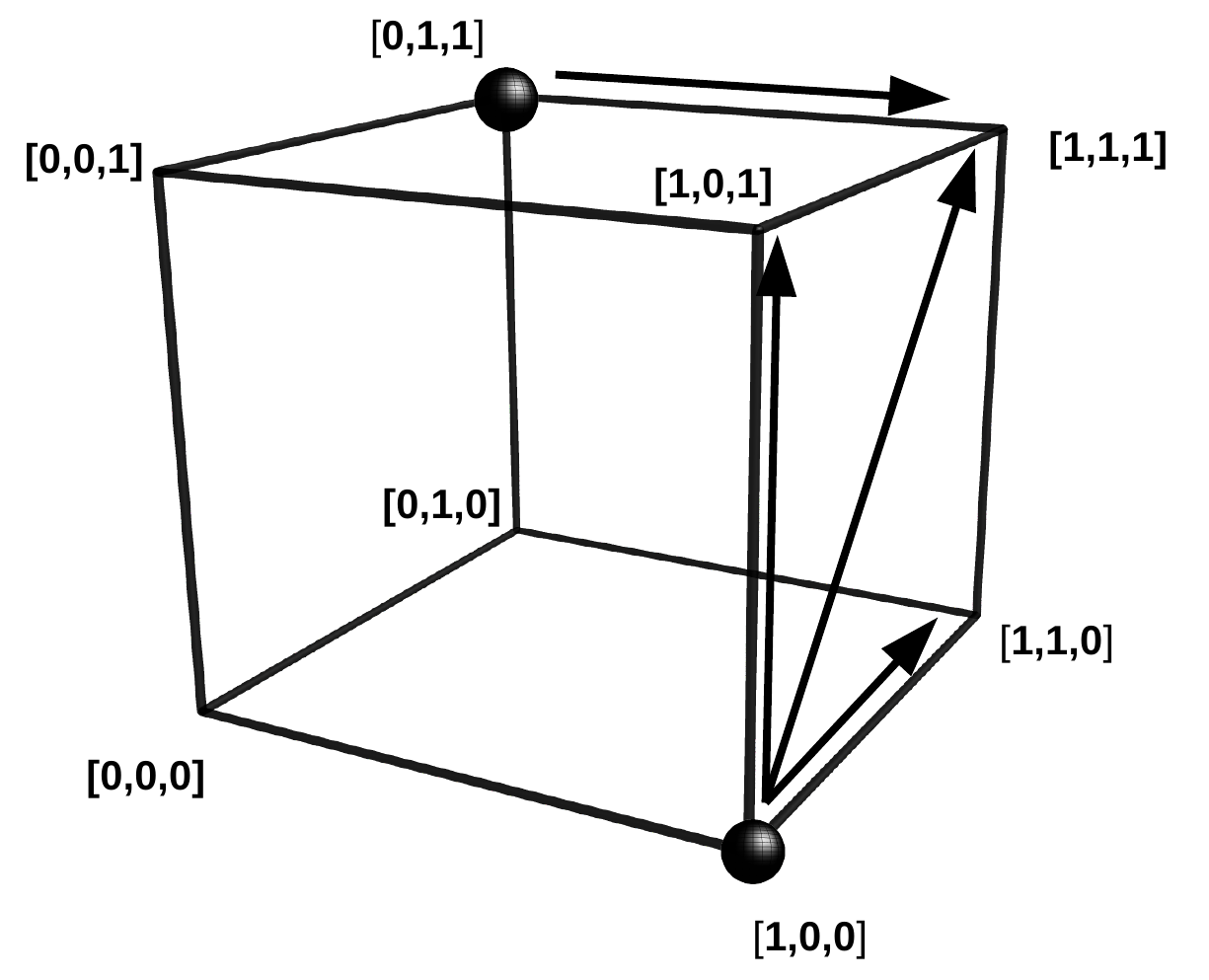} & \includegraphics[width=0.22\textwidth,trim={1cm 10.5cm 14.5cm 1cm},clip]{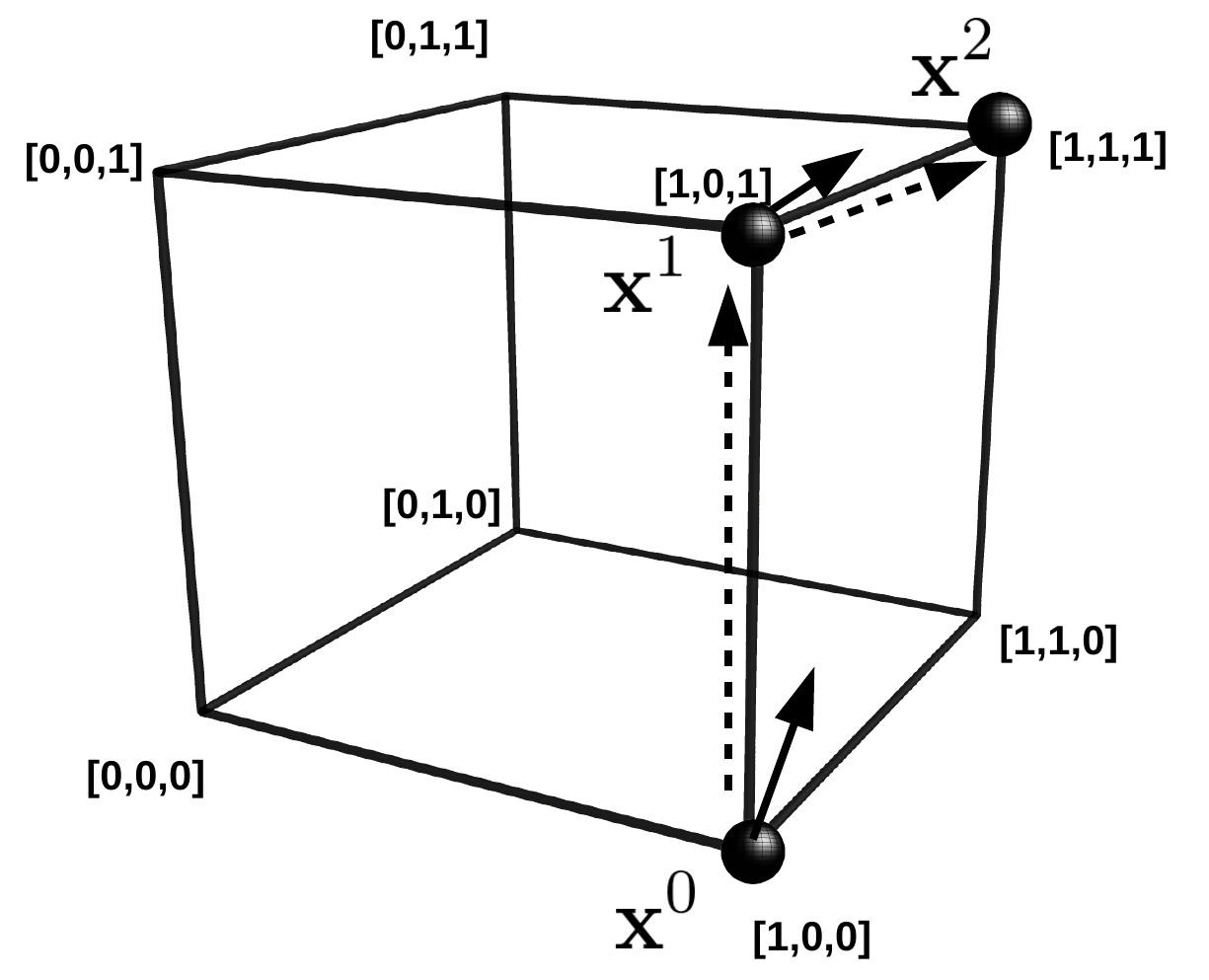}\\
		(a) & (b) \\
	\end{tabular}
	\caption{(a) Two malicious binary executables (malwares) in the 3-dimensional binary indicator vector space. The set of adversarial malware versions for the malware at $[1,0,0]$ is   $\mathcal{S}([1,0,0])=\{[1,0,0],[1,1,0],[1,0,1],[1,1,1]\}$, and for the malware at $[0,1,1]$ is $\mathcal{S}([0,1,1])=\{[0,1,1],[1,1,1]\}$. The arrows point to the set of allowed perturbations. (b) Two-step bit gradient ascent ($\texttt{BGA}^2$). The solid arrows represent the loss gradient at the arrows end points, while the dashed arrows represent the bit gradient ascent update step. At step 1,  the contribution to the magnitude of the loss gradient ($\ell_2$-norm) is predominantly towards setting the $3$rd feature. Thus, $\vx_1$ is obtained by setting $\vx_0$'s $3$rd bit. Similarly, $\vx_2$ is obtained by setting $\vx_1$'s $2$nd bit. After visiting 2 vertices besides our starting vertex, we choose  $\arg\max_{\vx\in \{\vx^0,\vx^1,\vx^2\}} L(\vtheta,\vx, 1)$ as the adversarial malware version of  $\vx^0$. Note that this a special case of \mBGA, where only one bit is set at a step. Two-step bit coordinate ascent ($\texttt{BCA}^2$) would generate the same adversarial malware version.}
	\label{fig:syn-example}
\end{figure}

Our malware-suited methods step off from the empirical success of gradient-based methods like $\fgsm^k$ in approximating inner maximizers for continuous feature spaces~\cite{madry2017towards}. The bulk of prior work has focused on adversarial attacks against images. In such setups, pixel-level perturbations are often constrained to $\ell_\infty$-ball around the image at hand~\cite{goodfellow2014explaining}. In the case of malware, perturbations that preserve malicious functionality  correspond to setting unset bits in the binary feature vector $\vx$ of the malware at hand. As depicted in Fig.~\ref{fig:syn-example} (a), we can only add features that are not present in the binary executable and never remove those otherwise. Thus, $\mathcal{S}(\vx)=\big\{\bar{\vx} \in \binaryset^m\;|\; \vx \wedge \bar{\vx} = \vx\big\}$ and $|\mathcal{S}(\vx)|= 2^{m-\vx^T\mathbf{1}}$. One could incorporate all the adversarial malware versions in the training through brute force enumeration but they grow exponentially in number and blind spots could be redundantly visited. On the other hand, with gradient-based methods, we aim to introduce adversarial malware versions in an online manner based on their difficulty in terms of model accuracy.

 \def\arraystretch{1.5}
 \begin{table*}
 	\caption{ \rm Proposed inner maximizers for the saddle-point problem~\eqref{eq:saddle-problem}.} 
 	\begin{center}
 				\resizebox{\textwidth}{!}{
 			\begin{tabular}{p{19.5cm}}
 				Considered inner maximization methods for crafting adversarial versions of a malware given its binary indicator vector $\vx$. Denote  ${\partial L(\vtheta, \vx^t, y)}/{\partial x^t_j}$ by $\shortdLdxj$, the adversarial malware version by $\bar{\vx}^{k}$,  the projection operator into the interval $[a,b]$ by $\Pi_{[a,b]}$ such that $\Pi_{[a,b]}=\max(\min(x,b),a)$, the OR operator by $\vee$, and the XOR operator by $\oplus$. Furthermore, for all the methods, the initial starting point $\vx^0$ can be any point from $\mathcal{S}(\vx)$, i.e., $\vx^0 \in \mathcal{S}(\vx)$. In our setup, $\vx^0$ is set to $\vx$. For \mBGA~and \mBCA, the $(1-2x^t_j)$ term is used to enforce that the gradient is towards $0$ if $x^t_j=1$, and vice versa. \\
 		\end{tabular}}
 		\resizebox{\textwidth}{!}{
 			\begin{tabular}{ll}
 				\toprule
 				\textbf{Method} & \textbf{Definition} \\
 				\toprule
 				\multirow{3}{*}{\mfgsm~with deterministic rounding (\dmfgsm)} &  $x^{t+1}_j =  \Pi_{[0,1]} \big(x^t_j + \epsilon \sgn(\shortdLdxj)\big) \;,\; 0 \leq j< m\;,\; 0 \leq t< k$\\
 				& $\bar{x}^{k}_j= \mathbf{1}\bigg\{ x^k_j > \alpha  \bigg\}\vee x_j\;,\; \alpha \in [0,1]\;,\; 0 \leq j< m$\\
 				& $\vx_{adv} \in \arg\max \{ L(\vtheta,  \vx^*, 1)\;|\;  \vx^* \in \{\bar{\vx}^k ,\vx\} \}$\\
 				\midrule 
 				\multirow{3}{*}{\mfgsm~with randomized rounding (\rmfgsm)} & $x^{t+1}_j =  \Pi_{[0,1]} \big(x^t_j + \epsilon \sgn(\shortdLdxj)\big) \;,\; 0 \leq j< m\;,\; 0 \leq t< k$\\
 				& $\bar{x}^{k}_j= \mathbf{1}\bigg\{ x^k_j > \alpha_j  \bigg\}\vee x_j\;,\; \alpha_j \in \mathcal{U}(0,1)\;,\; 0 \leq j< m$\\
 				& $\vx_{adv} \in \arg\max \{ L(\vtheta,  \vx^*, 1)\;|\;  \vx^* \in \{\bar{\vx}^k ,\vx\} \}$\\
 				\midrule 
 				\multirow{2}{*}{Multi-Step Bit Gradient Ascent  (\mBGA)} & $x^{t+1}_j =\bigg(x^{t}_{j} \oplus  \mathbf{1}\bigg\{ (1-2x^t_j)\;\shortdLdxj \geq \frac{1}{\sqrt{m}} ||\dLdx||_2  \bigg\}\bigg)\vee x_j\;,\; 0 \leq j< m\;,\; 0\leq t< k$ \\
 				&$\vx_{adv} \in \arg\max \{ L(\vtheta, \vx^*, 1)\;|\; \vx^* \in \{\vx^t\}_{ 0\leq t \leq k} \cup \{\vx\} \}$ \\
 				\midrule
 				\multirow{3}{*}{Multi-Step Bit Coordinate Ascent (\mBCA)} & 
 				$j^{t+1}\in \arg\max_{1\leq j\leq m} (1-2x^t_j)\;\shortdLdxj$\\ 
 				&	$x^{t+1}_{j}= (x^{t}_{j} \oplus \mathbf{1}\{j = j^{t+1}\} )\vee x_{j}\;,\; 0 \leq j< m\;,\; 0\leq t< k$ 
 				\\
 				& $\vx_{adv} \in \arg\max \{ L(\vtheta, \vx^*, 1)\;|\;  \vx^* \in \{\vx^t\}_{ 0\leq t \leq k} \cup \{\vx\} \}$\\
 				\bottomrule
 		\end{tabular}}
 	\end{center}
 	\label{tbl:maximizer-methods}	
 \end{table*}

In the continuous space of images, projected gradient descent can be used to incorporate the $\ell_\infty$-ball constraint (e.g., the \texttt{Clip} operator~\cite{kurakin2016adversarial}). Inspired by linear programming relaxation and rounding schemes for integer problems, we extend the projection operator to make use of gradient-based methods for the malware binary space via \textbf{deterministic} or \textbf{randomized} \textbf{rounding} giving rise to two discrete, binary-encoded, constraint-based variants of \mfgsm, namely\textbf{ \dmfgsm} and\textbf{ \rmfgsm}, respectively.  It is interesting to note that \malgan's black-box system~\cite{hu2017generating} used a deterministic rounding scheme---with $\alpha=0.5$---to craft adversarial malware versions.  

With \mfgsm~in continuous space, \AEs are generated by moving iteratively in the feasible space (e.g., the $\ell_\infty$-ball around around an image). In contrast, the crafted adversarial malware versions are situated at the vertices  of the binary feature space. Instead of multi-stepping through the continuous space to generate just one adversarial malware version (i.e., \dmfgsm~or \rmfgsm), we can use the gradient to visit multiple feasible vertices (i.e., adversarial malware versions) and choose the one with the maximum loss, see Fig.~\ref{fig:syn-example}~(a). This suggests a third method: \textbf{multi-step Bit Gradient Ascent (\mBGA)}, see Fig.~\ref{fig:syn-example}~(b). This method sets the bit of the $j$th feature if the corresponding partial derivative of the loss is greater than or equal to the loss gradient's $\ell_2$-norm divided by $\sqrt{m}$. The rationale behind this is that the projection of a unit vector with equal components onto any coordinate equals  $1/\sqrt{m}$. Therefore we set bits (features) whose corresponding partial derivative contribute more or equally to the $\ell_2$-norm of the gradient in comparison to the rest of the features. After $k$ steps, the binary indicator vector that corresponds to the vertex with the maximum loss among the visited vertices is chosen as the adversarial malware version. A final method: \textbf{multi-step Bit Coordinate Ascent (\mBCA)} updates one bit in each step by considering the feature with the maximum corresponding partial derivative of the loss.  A similar approach has been shown effective for Android malware evasion in~\cite{grosse2016adversarial,grosse2017adversarial}. Table~\ref{tbl:maximizer-methods} presents a formal definition of the methods. In the next section, we propose a metric to measure their effectiveness in covering the model's blind spots.

\subsection{Blind Spots Coverage} With adversarial learning, we aim to discover and address blind spots of the model while learning its parameters simultaneously. In other words, we would like to incorporate as many members of $\mathcal{S}^*(\vx)$ as possible in training the model. In line with this notion, we propose a new measure called the \textbf{blind spots covering number}, denoted $\bscn$,  which measures the effectiveness of an algorithm $\mathcal{A}$ in computing the inner maximizers of~\eqref{eq:saddle-problem}. The measure is defined as the expected ratio of the number of adversarial malware versions crafted by $\mathcal{A}$ during training, denoted by $\mathcal{S}^*_{\mathcal{A}}(\vx)$, to the maximum possible number of the same. Formally, it can be written as follows.

\begin{equation}
\bscn(\mathcal{A}) = \expect_{(\vx,y)\sim\distrib}\bigg[ \frac{y |\mathcal{S}_{\mathcal{A}}^*(\vx)|}{2^{m-\vx^T\mathbf{1}}} \bigg]
\label{eq:bscn}
\end{equation}

Models trained with high $\bscn$ have seen more \AEs in training, and because training against multiple \AEs implies more exhaustive approximations of the inner maximization problem, they are expected to be more robust against adversarial attacks~\cite{madry2017towards}. While it may be computationally expensive to compute~\eqref{eq:bscn} exactly, we provide a probabilistic approximation of it in \cref{sec:results}.

\subsection{Adversarial Learning Framework}

\begin{figure}[tb]
	\centering
	\includegraphics[width=0.49\textwidth]{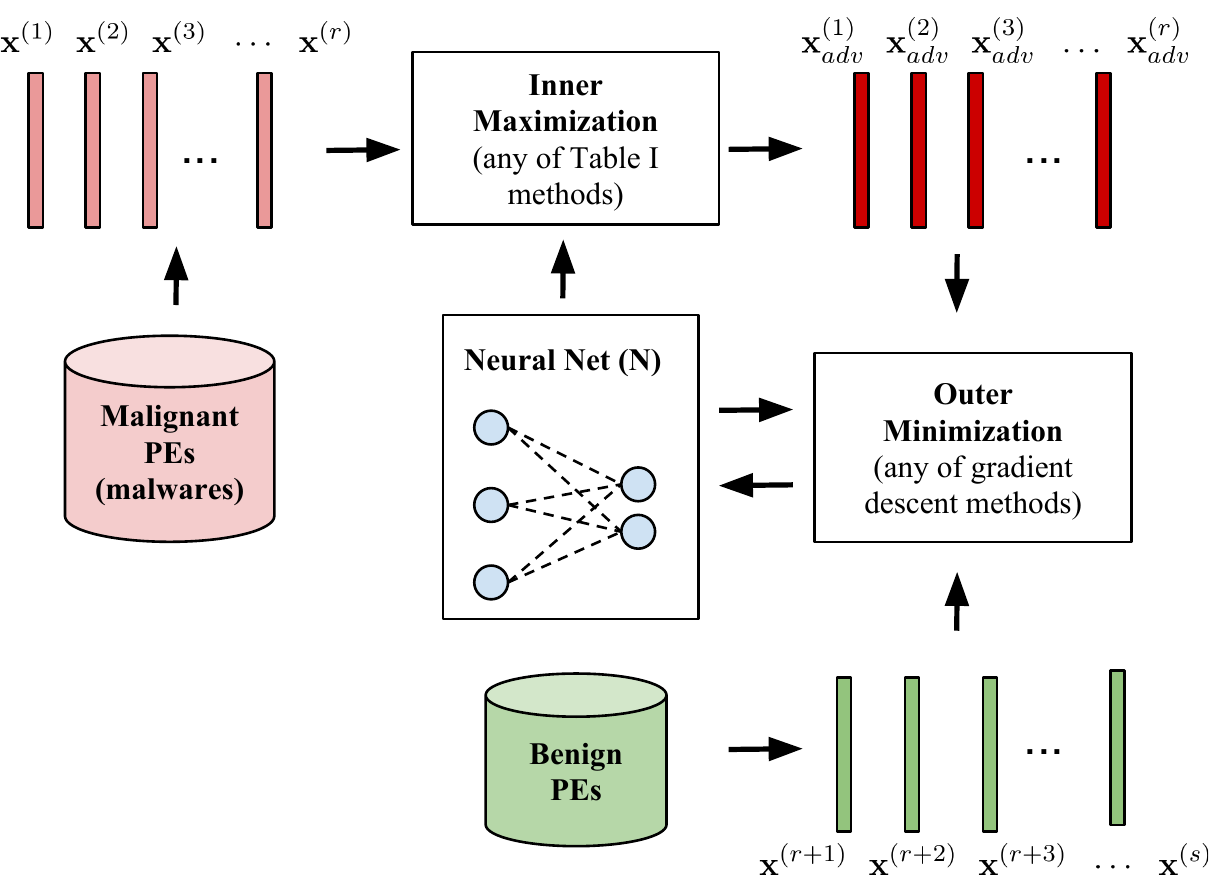}

	\caption{Overview of the \horse framework. Malware is perturbed by an inner maximization method (any of Table~\ref{tbl:maximizer-methods}) to create \AEs. The generated adversarial malware versions and benign examples are used in an outer minimization of the adversarial and natural loss~\eqref{eq:saddle-problem-empirical}, which can be solved in minibatches using any variant of the gradient descent algorithm.}
	\label{fig:RAMOverview}
\end{figure}

Having specified four methods for approximating the inner maximizers of~\eqref{eq:saddle-problem} and a measure of their effectiveness, 
we can now describe \horse, our adversarial learning framework for robust malware detection. Consider a training dataset $D$ of $n$ independent and identically distributed samples drawn from $\distrib$. As outlined in Algorithm~\ref{alg:adv-training} and depicted in Fig.~\ref{fig:RAMOverview}, \horse groups $D$ into minibatches $B$ of  $s$ examples similar to~\cite{kurakin2016adversarial}. However, the grouping here is governed by the examples' labels: the first $r<s$ examples are malicious, followed by $s-r$ benign examples. At each training step, the model's parameters $\vtheta$ are optimized with respect to the adversarial loss~\eqref{eq:adv-samples} of malware executables and the natural loss~\eqref{eq:risk} of benign executables. This is motivated by the fact that authors of benign applications have no interest in having their binaries misclassified as malwares~\cite{grosse2016adversarial}.  However, one should note that a malware author might wish to create adversarial benign applications to poison the training dataset. This possibility is considered for future work. As an equation, our empirical saddle-point problem at each training step has the form
	\begin{equation}
\min_{\vtheta \in \mathbb{R}^p} \frac{1}{s}\bigg[\sum_{i=1}^{r} \max_{\bar{\vx}^{(i)} \in \mathcal{S}(\vx^{(i)})} L(\vtheta, \bar{\vx}^{(i)}, 1)+ \sum_{i=r+1}^{s} L(\vtheta, \vx^{(i)}, 0)\bigg]\;.
\label{eq:saddle-problem-empirical}
\end{equation}

\begin{algorithm}
	\small
	\caption{\horse \newline
		\textbf{Input:}\newline 
		$~$\hspace{\algorithmicindent}$N$ : neural network model, 
		$D$ : training dataset, \newline
		$~$\hspace{\algorithmicindent}$s$ : minibatch size, 
		$r$ : number of malwares in minibatch, \newline
		$~$\hspace{\algorithmicindent}$\mathcal{A}$ : inner maximizer algorithm (any of Table~\ref{tbl:maximizer-methods}) 
	}\label{alg:adv-training}
	\begin{algorithmic}[1]
		\State Randomly initialize network $N$
		\Repeat
		\State Read minibatch $B$ from dataset $D$ 
		 $$B = \{ \vx^{(1)}, \ldots, \vx^{(s)}\,|\; y^{i\leq r}=1, \;y^{i>r}=0 \}$$
		\State Generate $r$ adversarial versions $\{ \vx^{(1)}_{adv}, \ldots, \vx^{(r)}_{adv} \}$ 
		\Statex \hspace{\algorithmicindent}from feasible sets of corresponding malware examples  
		\Statex \hspace{\algorithmicindent}$\{ \mathcal{S}(\vx^{(1)}), \ldots, \mathcal{S}(\vx^{(r)}) \}$ by $\mathcal{A}$ using current state  of $N$ 
		\State Make new minibatch $$B'=\{ \vx^{(1)}_{adv}, \ldots, \vx^{(r)}_{adv}, \vx^{(r+1)}, \ldots, \vx^{(s)} \}$$
		\State Do one training step of network $N$ with minibatch $B'$
		\Until training converged
	\end{algorithmic}
\end{algorithm}

\section{Experiments}
\label{sec:experiments}

This section provides an empirical evaluation of our proposition in \cref{sec:method}. We conduct experiments to validate and compare the efficacy of the proposed methods in terms of classification accuracy, evasion rates, and blind spots coverage. First, the setup of our experiments is described in \cref{sec:setup}, followed by a presentation of the results in \cref{sec:results}.


\subsection{Setup}
\label{sec:setup}

\textbf{Dataset.} The Portable Executable~(PE)
format\cite{peformat}
is a file format for executables in Windows operating systems. The
format encapsulates information necessary for Windows OS to manage the
wrapped code. PE files have widespread use as malware. We created a corpus of malicious and benign PE files from VirusShare\cite{virusshare} and internet download sites, respectively. 

To label the collected PEs,  we use VirusTotal's\cite{virustotal} ensemble of
virus detectors. We require benign files to have 0\% positive
detections from the ensemble and malicious files to have greater than
50\% positive detections to avoid false positives. At the time of writing this paper, we have 34,995
malicious and 19,696 benign PEs.

\textbf{Feature Representation.} As mentioned earlier, each portable executable is represented as a binary indicator feature
vector. Each index of the feature vector represents a unique Windows
API call and a "1" in a location represents the presence of the
corresponding API call. In our dataset of PEs, we found a total of
22,761 unique API calls. Thus, each PE file is represented by a binary indicator vector~$\vx \in \binaryset^m$, with $m=22,761$. We use the \texttt{LIEF}\cite{lief} library
 to parse each PE and
turn it into its representative binary feature vector.\footnote{The generated feature vectors are available by request.}

\textbf{Neural Net ($N$) Architecture.} We use a feed-forward
network for our malware classifier $N$ with 3 hidden layers of 300 neurons each. The ReLU activation function is applied to all the $3\times300$ hidden neurons. The LogSoftMax function is applied to the output layer's two neurons which correspond to the two labels at hand: benign and malicious. The model is implemented in \texttt{PyTorch}~\cite{paszke2017automatic}.

\textbf{Learning Setup.} We use $19,000$ benign PEs and $19,000$ malicious PEs to construct our training ($60\%$), validation ($20\%$), and test ($20\%$) sets. The training set is grouped into minibatches of $16$ PE samples according to Line~3 of Algorithm~\ref{alg:adv-training}. The classifier $N$'s parameters $\vtheta$ are tuned with respect to~\eqref{eq:saddle-problem-empirical}, where $L$ is the negative log likelihood loss, using the \texttt{ADAM} optimization algorithm with a 0.001 learning rate over 150 epochs. Note that one step of \texttt{ADAM} corresponds to Line 6 of Algorithm~\ref{alg:adv-training}. To avoid overfitting, model parameters at the minimum validation loss are used as the final learned parameters $\vtheta^*$.
With regard to the inner maximizers algorithms (Table~\ref{tbl:maximizer-methods}), all were set to perform $50$ steps, i.e., $k=50$. This makes the step size~$\epsilon$ for \dmfgsm~and \rmfgsm~ small enough (we set it to $\epsilon=0.02$) to follow the gradient accurately while also ensuring that multi-steps could reach close to other vertices of the binary feature space (Fig.~\ref{fig:syn-example}) and not be suppressed by rounding. With $50$ steps and $0.02$ step size, both these conditions are met. We run Algorithm~\ref{alg:adv-training} with $\mathcal{A}$ being set to each of the inner maximizers from Table~\ref{tbl:maximizer-methods} to obtain 4 adversarially trained models in addition to the model trained naturally. We also used the adversarial sample crafting method presented by Grosse \textit{et al.}~\cite[Algorithm 1]{grosse2017adversarial} which trains a model adversarially without using a saddle-point formulation: the \AEs in~\cite{grosse2017adversarial} are tuned with respect to the value of the benign output neuron rather than the loss $L$. Though not directly, this does maximize the adversarial loss value.  All experiments for the six models were run on a CUDA-enabled GTX 1080 Ti GPU.

\subsection{Results}
\label{sec:results}
For brevity, we refer to the trained models by their inner maximizer methods. Experiment results are presented in Tables~\ref{tbl:accuracy} and~\ref{tbl:evasion-rates} as follows.

\textbf{Classification Performance.} Based on Table~\ref{tbl:accuracy},  all the adversarially trained models achieve a classification accuracy comparable  to the naturally trained counterpart. However, 
we observe that models trained using inner maximizers of Table~\ref{tbl:maximizer-methods} tend to have higher false positive rate (FPR) and lower false negative rate (FNR)---positive denotes malicious. The FPR increase can be explained by the transforming of malware samples when models are trained adversarially. Such transformations could turn malware feature vectors into ones more similar to those of the benign population, and subsequently the benign test set. Likewise, the FNR decrease can be attributed to the adversarial versions boosting the model's confidence on vertices with less original malicious samples compared to the benign samples. With~\cite{grosse2017adversarial}'s method, it is the other way around.  Arguably, the reason is that its adversarial objective is to maximize just the benign (negative) neuron's output and it is indifferent to the malicious (positive) neuron. As a result, the crafted adversarial malware version does not necessarily end up at a vertex at which the model's confidence, with respect to the malicious label, is low, which consequently improves the FPR and worsens the FNR.  

\textbf{Robustness to Evasion Attacks.} We tried the adversarial attackers generated by the inner maximizers and~\cite{grosse2017adversarial}'s method as  inputs to each of the trained models to assess their robustness against the adversaries generated during training as well as other adversaries. It can be seen in Table~\ref{tbl:evasion-rates} that \rmfgsm~is our most successful adversarial training method, achieving relatively low evasion rates across all attack methods. As expected, all training methods are resistant to attacks using the same method, but each method aside from \rmfgsm~has at least one adversarial method that it performs poorly against. Evasion rates for \texttt{Natural} training, which uses non-altered malicious samples, provide a baseline for comparison.

\textbf{Blind Spots Coverage.} Given the high-dimension feature vectors and the sizeable dataset, it was computationally expensive to compute $\bscn$ exactly. Instead, we computed an approximate probabilistic measure $\bar{\mathcal{N}}_{BS}$ using a Bloom filter~\cite{bloom1970space}. The computed measures are presented in the last column of Table~\ref{tbl:accuracy} as the ratio of total adversarial malware versions to original samples over all the training epochs. \texttt{Natural} training has a ratio of $1.0$ since we do not modify the malicious samples in any way. A coverage value of $4.0$ for \rmfgsm~means that with \textit{high probability} we explored $4$ times as many malicious samples compared to \texttt{Natural} training. A high coverage value indicates that the adversarial training explored more of the valid region $\mathcal{S}(\vx)$ for malware sample $\vx$, resulting in a more robust model. This observation is substantiated by the correlation between coverage values in Table~\ref{tbl:accuracy} and evasion rates in Table~\ref{tbl:evasion-rates}. Note that $\bar{\mathcal{N}}_{BS}$ is computed and updated after each training step. Thus, it can be used as an online measure to assess  training methods' robustness to adversarial attacks.

\begin{table}[t]
	\centering
	\caption{ Performance metrics of the trained models.} 
	\label{tbl:accuracy}
	\resizebox{0.49\textwidth}{!}{
	\begin{tabular}{p{10cm}}
		In percentage, \textbf{Accuracy}, False Positive Rate (\textbf{FPR}), and False Negative Rate (\textbf{FNR}) are of the test set: 3800 malicious PEs and 3800 bengin PEs, with $k=50$. $\bar{\mathcal{N}}_{BS} $ denotes the probabilistic normalized measure  computed during training to approximate the blind spots covering number $\mathcal{N}_{BS}$. This was obtained using a Bloom filter to track the number of distinct malware samples presented during training, be they from the original malware  training samples or their adversarial versions. Models corresponding to bold cells are the best with regard to the corresponding measure/rate. The measures of the inner maximizers and~\cite{grosse2017adversarial}'s are reported in their relative difference to \texttt{Natural}'s counterparts.\\
	\end{tabular}}
	\resizebox{0.35\textwidth}{!}{
		\begin{tabular}{lcccccc}
			\toprule
			\textbf{Model} & \textbf{Accuracy} & \multicolumn{1}{c}{\textbf{FPR}} & \multicolumn{1}{c}{\textbf{FNR}} & $\mathbf{\bar{\mathcal{N}}_{BS}} $\\
			\midrule
			\bf \texttt{Natural}&      $91.9$ &  \multicolumn{1}{c}{$8.2$} &  \multicolumn{1}{c}{$8.1$} & $1.0$ \\
			\midrule
			\bf \dmfgsm &      $+0.1$ &   {${+1.4}$} &   { ${-1.7}$}& $+1.6$\\
			\bf \rmfgsm &      $-0.6$ & { ${+3.6}$}&   { $\mathbf{-2.4}$}& $\mathbf{+3.0}$\\
			\bf \mBGA  &      $\mathbf{+0.2}$ &   { ${+0.0}$}&  { ${-0.5}$}& $+2.5$\\
			\bf \mBCA   &      $-0.3$ &   { ${+0.9}$}&  { ${-0.5}$}& $+0.0$\\
			\midrule
			\bf \hspace{1sp}\cite{grosse2017adversarial}'s method  &      $-1.1$ &   { $\mathbf{-3.9}$}&   { ${+5.9}$}& $+0.6$\\
			\bottomrule
	\end{tabular}}
\end{table}

\begin{table}[t]
	\centering
	\caption{Evasion Rates.}
	\label{tbl:evasion-rates}
		\resizebox{0.49\textwidth}{!}{
		\begin{tabular}{p{10cm}}
			Evasion rates of adversaries on the test set against the trained models with $k=50$. Models corresponding to bold cells are the most robust models with regard to the corresponding adversary. Adversaries corresponding to shaded cells are  the (or one of the) most successful adversaries with regard to the corresponding model. Evasion rates of the proposed inner maximizers are the lowest on their corresponding expected adversary after the \texttt{Natural} adversary as shown by the corresponding framed cells along the diagonal. This conforms to their saddle-point formulation, in contrast to \cite{grosse2017adversarial}'s method with \mBCA~being its weakest adversary after the \texttt{Nautral} adversary, as framed below. This is expected as training with~\cite{grosse2017adversarial}'s method does not follow an exact saddle-point formulation. \\
	\end{tabular}}
	\resizebox{0.48\textwidth}{!}{
		\begin{tabular}{lcccccc}
			\toprule
			\multirow{2}{*}{
				\textbf{\hspace{1em}Model}} & \multicolumn{6}{c}{\textbf{Adversary}}\\
			\cmidrule{2-7}
			&  \bf \texttt{Natural} &    \bf \dmfgsm &    \bf \rmfgsm &    \bf   \mBGA &      \bf \mBCA & \bf \hspace{1sp}\cite{grosse2017adversarial}'s method \\
			\midrule
			\bf \texttt{Natural}&      $8.1$ &    \cfbox{$99.7$} &     \cfbox{$99.7$} &   \cfbox{$99.7$} &   $41.7$ &    \cfbox{$99.7$} \\
			\midrule
			\bf \dmfgsm &      $6.4$ &  \fbox{$\mathbf{6.4}$} &     $21.1$ &    $7.3$ &   $27.4$ &    \cfbox{$99.2$} \\
			\bf \rmfgsm &      $\mathbf{5.7}$ &      $7.0$ &      \fbox{{$\mathbf{5.9}$}} &    \textbf{5.9} &   
			\textbf{6.8} &    \cfbox{$35.0$} \\
			\bf \mBGA   &      $7.6$ &     $39.6$ &     $17.8$ &    \fbox{$7.6$} &   $10.9$ &    \cfbox{$68.4$} \\
			\bf \mBCA   &      $7.6$ &     \cfbox{$99.5$} &     \cfbox{$99.5$} &   $91.8$ &    \fbox{$7.9$} &    $98.6$ \\
			\midrule
			\bf \hspace{1sp}\cite{grosse2017adversarial} 's method &     $14.0$ &     \cfbox{$69.3$} &     \cfbox{$69.3$} &   $37.5$ &   \fbox{$14.1$} &    $\mathbf{15.6}$ \\
			\bottomrule
	\end{tabular}}
\end{table}

\section{Conclusions and Future Work}
\label{sec:concl-future-wrok}

We investigated methods that reduce the adversarial blind spots for neural network
malware detectors. We approached this as a saddle-point optimization problem in the
binary domain and used this to train DNNs via multiple inner
maximization methods that are robust to adversarial malware
versions of the dataset. 

We used a dataset of PE files to assess the robustness against
evasion attacks. Our experiments have demonstrated once again the power of randomization in addressing challenging problems, conforming to the conclusions provided by state-of-art attack papers~\cite{carlini2017adversarial}. Equipping projected gradient descent with randomness in rounding helped uncover roughly 4 times as many malicious samples in the binary feature space as those uncovered in natural training. This performance correlated with the online measure we introduced to assess the general expectation of robustness.

There are several future research questions. First, we would like to study the loss landscape of the adversarial malware versions and the effect of starting point $\vx^0$  initialization for inner maximizers, in comparison to their continuous-domain counterparts. Second,  $\bscn$ quantifies how many different adversarial examples are generated but it does not capture how they are located with regard to the benign examples and, subsequently, their effect on the model's FPR and FNR. We hope that investigating these directions will lead towards fully resistant deep learning models for malware detection.

\section*{Acknowledgment}

This work was supported by the MIT-IBM Watson AI Lab and CSAIL CyberSecurity Initiative.

\bibliographystyle{plain}
\bibliography{bibliography}

\end{document}